# Relativistic new Yukawa-like potential and tensor coupling

Sameer M. Ikhdair[1], Majid Hamzavi[2*]

[1]*Physics Department, Near East University, 922022 Nicosia, North Cyprus, Mersin 10, Turkey*

[2] *Department of Basic Sciences, Shahrood Branch, Islamic Azad University, Shahrood, Iran*

[*]*Corresponding author: Tel.:+98 273 3395270, fax: +98 273 3395270*

[2] Email: majid.hamzavi@gmail.com

[1] Email: sikhdair@neu.edu.tr

**Abstract**

We approximately solve the Dirac equation for a new suggested generalized inversely quadratic Yukawa (GIQY) potential including a Coulomb-like tensor interaction with arbitrary spin-orbit coupling quantum number $\kappa$. In the framework of the spin and pseudospin (p-spin) symmetry, we obtain the energy eigenvalue equation and the corresponding eigenfunctions, in closed form, by using the parametric Nikiforov-Uvarov (NU) method. The numerical results show that the Coulomb-like tensor interaction, $-T/r$, removes degeneracies between spin and p-spin state doublets. The Dirac solutions in the presence of exact spin symmetry are reduced to Schrödinger solutions for Yukawa and inversely quadratic Yukawa potentials.

## 1. Introduction

Relativistic symmetries of the Dirac Hamiltonian had been discovered many years ago. However, these symmetries have been recently recognized empirically in nuclear and hadronic spectroscopies [1]. Within the framework of Dirac equation, the pspin symmetry used to feature deformed nuclei and the superdeformation to establish an effective shell-model [2-4]. The spin symmetry is relevant to mesons [5] and occurs when the difference of the scalar $S(r)$ and vector $V(r)$ potentials are constant, i.e., $\Delta(r) = C_s$ and the p-spin symmetry occurs when the sum of the scalar and vector potentials are constant, i.e., $\Sigma(r) = C_{ps}$ [6-7]. The p-spin symmetry refers to a quasi-degeneracy of single nucleon doublets with non-relativistic quantum number $(n, l, j = l + 1/2)$ and $(n-1, l+2, j = l+3/2)$, where $n$, $l$ and $j$ denote the single nucleon radial, orbital and total angular quantum numbers, respectively [8-9]. Further, the total angular momentum is $j = \tilde{l} + \tilde{s}$, where $\tilde{l} = l+1$ pseudo-angular momentum and $\tilde{s}$ is p-spin angular momentum [10]. Recently, the tensor potentials were introduced into the Dirac equation with the substitution $\vec{p} \to \vec{p} - im\omega\beta.\hat{r}U(r)$ and a spin-orbit coupling is added to the Dirac Hamiltonian [11-12]. Lisboa et al [13] have studied a generalized relativistic harmonic oscillator for spin-1/2 particles by considering a Dirac Hamiltonian that contains quadratic vector and scalar potentials together with a linear tensor potential under the conditions of p-spin and spin symmetry. Alberto et al [14] studied the contribution of the isoscalar tensor coupling to the realization of p-spin symmetry in nuclei. Akçay [15] solved exactly the Dirac equation with scalar and vector quadratic potentials including a Coulomb-like tensor potential. He also solved exactly the Dirac equation for a linear and Coulomb-like term containing the tensor potential too [16]. Aydoğdu and Sever [17] obtained exact solution of the Dirac equation for the pseudoharmonic potential in the presence of linear tensor potential under the p-spin symmetry and showed that tensor interactions remove all degeneracies between members of p-spin doublets. Ikhdair and Sever [10] solved approximately Dirac-Hulthén problem under spin and p-spin symmetric limits including Coulomb-like tensor potential with arbitrary spin-orbit coupling number $\kappa$. Very recently, Hamzavi et al [18,19] presented exact solutions of the Dirac equation for Mie-type potential and approximate solutions of the Dirac-Morse problem with a Coulomb-like tensor potential. Very recently, Hamzavi et al [20] introduced a novel potential they called inversely quadratic Yukawa (IQY) potential and solved Dirac-IQY problem in the presence of spin and p-spin symmetric limits with Coulomb-like tensor interaction. Over the past few years, the Schrödinger, Klein-Gordon (KG) and the Dirac equations have been solved for various types of quantum potential models by different authors [21-40].

In this work, we introduce a novel potential for the first time we call it the generalized inversely quadratic Yukawa (GIQY) potential having similar behavior as the IQY potential [20] and Yukawa potential [40] at the entire positive line range, $r \in (0, \infty)$ as shown in Figure 1. In the short screening regime, when taking weak coupling strength $V_{\text{GIQY}}(r)$ is close to $V_{\text{IQY}}(r)$ as shown in Figure 1a, however, when taking strong coupling strength $V_{\text{GIQY}}(r)$ is close to $V_{\text{Y}}(r)$ as seen in Figure 1b. It has the form:

$$V(r) = -V_0 \left(1 - \frac{1}{r} e^{-\alpha r}\right)^2 = -\frac{A}{r^2} e^{-2\alpha r} + \frac{B}{r} e^{-\alpha r} - C, \; A = C = V_0, \; B = 2V_0, \quad (1)$$

where $\alpha$ is the screening parameter and $V_0$ is the coupling strength of the potential. The above potential is simply a superposition of two potentials; namely, the IQY-plus-Yukawa-like potential. It should come asymptotically to a finite value as $r \to \infty$ and should become infinite at $r = 0$. A form of Yukawa potential has been earlier used by Taseli [41] in obtaining modified Laguerre basis for hydrogen-like systems. Also, Kermode et al [42] have used different forms of the Yukawa potential to obtain the effective range functions. But, not much has been done in solving the Varshni potential.

The motivation of the present work is to study and test a more generalized form of the Yukawa potential consisting of the IQY-plus-Yukawa-type potentials. We shall study the bound states of a spin-$1/2$ particle under the field of the GIQY potential (1) in view of spin and p-spin symmetries taking into account a Coulomb field interaction. We calculate the energy eigenvalues and the corresponding wave functions expressed in terms of Jacobi polynomials.

This paper is organized as follows. In Section 2, we briefly introduce the Dirac equation with scalar and vector potentials with arbitrary spin-orbit coupling quantum number $\kappa$ including tensor interaction under spin and p-spin symmetry limits. The parametric Nikiforov-Uvarov (NU) method is presented in Section 3. The energy eigenvalue formulas and corresponding wave functions are obtained in Section 4. The Schrödinger solutions for GIQY, IQY and Yukawa potentials are obtained from the exact spin symmetric limit by making appropriate transformations. We also make some remarks and present numerical results. Finally, we give our conclusion in Section 5.

**2. Dirac equation with tensor coupling potential**

The Dirac equation for fermionic massive spin-$1/2$ particles moving in the field of an attractive scalar $S(r)$, repulsive vector $V(r)$ and a tensor $U(r)$ potentials (in units $\hbar = c = 1$) is

$$[\vec{\alpha}.\vec{p} + \beta(M + S(r)) - i\beta\vec{\alpha}.\hat{r}U(r)]\psi(\vec{r}) = [E - V(r)]\psi(\vec{r}), \qquad (2)$$

where $E$ is the relativistic binding energy of the system, $\vec{p} = -i\vec{\nabla}$ is the three-dimensional momentum operator and $M$ is the mass of the fermionic particle. $\vec{\alpha}$ and $\beta$ are the $4\times 4$ Dirac matrices given by

$$\vec{\alpha} = \begin{pmatrix} 0 & \vec{\sigma} \\ \vec{\sigma} & 0 \end{pmatrix}, \quad \beta = \begin{pmatrix} I & 0 \\ 0 & -I \end{pmatrix}, \qquad (3)$$

where $I$ is $2\times 2$ unitary matrix and $\vec{\sigma}$ are three-vector spin matrices

$$\sigma_1 = \begin{pmatrix} 0 & 1 \\ 1 & 0 \end{pmatrix}, \quad \sigma_2 = \begin{pmatrix} 0 & -i \\ i & 0 \end{pmatrix}, \quad \sigma_3 = \begin{pmatrix} 1 & 0 \\ 0 & -1 \end{pmatrix}. \qquad (4)$$

The eigenvalues of spin-orbit coupling operator are $\kappa = \left(j + \frac{1}{2}\right) > 0$ and $\kappa = -\left(j + \frac{1}{2}\right) < 0$ for unaligned spin $j = l - \frac{1}{2}$ and the aligned spin $j = l + \frac{1}{2}$, respectively. The set $(H^2, K, J^2, J_z)$ can be taken as the complete set of the conservative quantities with $\vec{J}$ is the total angular momentum operator and $K = (\vec{\sigma}.\vec{L} + 1)$ is the spin-orbit where $\vec{L}$ is orbital angular momentum of the spherical nucleons that commutes with the Dirac Hamiltonian. Thus, the spinor wave functions can be classified according to their angular momentum $j$, the spin-orbit quantum number $\kappa$, and the radial quantum number $n$. Hence, they can be written as follows

$$\psi_{n\kappa}(\vec{r}) = \begin{pmatrix} f_{n\kappa}(\vec{r}) \\ g_{n\kappa}(\vec{r}) \end{pmatrix} = \frac{1}{r}\begin{pmatrix} F_{n\kappa}(r) Y^l_{jm}(\theta,\varphi) \\ iG_{n\kappa}(r) Y^{\tilde{l}}_{jm}(\theta,\varphi) \end{pmatrix}, \qquad (5)$$

where $f_{n\kappa}(\vec{r})$ is the upper (large) component and $g_{n\kappa}(\vec{r})$ is the lower (small) component of the Dirac spinors. $Y^l_{jm}(\theta,\varphi)$ and $Y^{\tilde{l}}_{jm}(\theta,\varphi)$ are spin and p-spin spherical harmonics, respectively, and $m$ is the projection of the angular momentum on the $z$-axis. Substituting Eq. (5) into Eq. (2) and making use of the following relations

$$(\vec{\sigma}.\vec{A})(\vec{\sigma}.\vec{B}) = \vec{A}.\vec{B} + i\vec{\sigma}.(\vec{A}\times\vec{B}), \qquad (6a)$$

$$\left(\vec{\sigma}.\vec{P}\right) = \vec{\sigma}.\hat{r}\left(\hat{r}.\vec{P} + i\frac{\vec{\sigma}.\vec{L}}{r}\right), \tag{6b}$$

together with the properties

$$\begin{aligned}
\left(\vec{\sigma}.\vec{L}\right)Y_{jm}^{\tilde{l}}(\theta,\phi) &= (\kappa-1)Y_{jm}^{\tilde{l}}(\theta,\phi), \\
\left(\vec{\sigma}.\vec{L}\right)Y_{jm}^{l}(\theta,\phi) &= -(\kappa-1)Y_{jm}^{l}(\theta,\phi), \\
\left(\vec{\sigma}.\hat{r}\right)Y_{jm}^{\tilde{l}}(\theta,\phi) &= -Y_{jm}^{l}(\theta,\phi), \\
\left(\vec{\sigma}.\hat{r}\right)Y_{jm}^{l}(\theta,\phi) &= -Y_{jm}^{\tilde{l}}(\theta,\phi),
\end{aligned} \tag{7}$$

one can obtain two coupled differential equations whose solutions are the lower $G_{n\kappa}(r)$ and upper $F_{n\kappa}(r)$ radial wave functions as

$$\left(\frac{d}{dr} + \frac{\kappa}{r} - U(r)\right)F_{n\kappa}(r) = \left(M + E_{n\kappa} - \Delta(r)\right)G_{n\kappa}(r), \tag{8a}$$

$$\left(\frac{d}{dr} - \frac{\kappa}{r} + U(r)\right)G_{n\kappa}(r) = \left(M - E_{n\kappa} + \Sigma(r)\right)F_{n\kappa}(r), \tag{8b}$$

where

$$\Delta(r) = V(r) - S(r), \tag{9a}$$

$$\Sigma(r) = V(r) + S(r). \tag{9b}$$

Eliminating $F_{n\kappa}(r)$ and $G_{n\kappa}(r)$ from Eqs. (8a) and (8b), we thus obtain two Schrödinger-like differential equations for the upper and lower radial spinor components

$$\begin{aligned}
&\left[\frac{d^2}{dr^2} - \frac{\kappa(\kappa+1)}{r^2} + \frac{2\kappa}{r}U(r) - \frac{dU(r)}{dr} - U^2(r)\right]F_{n\kappa}(r) \\
&+ \frac{\frac{d\Delta(r)}{dr}}{M + E_{n\kappa} - \Delta(r)}\left(\frac{d}{dr} + \frac{\kappa}{r} - U(r)\right)F_{n\kappa}(r) \\
&= \left[\left(M + E_{n\kappa} - \Delta(r)\right)\left(M - E_{n\kappa} + \Sigma(r)\right)\right]F_{n\kappa}(r),
\end{aligned} \tag{10}$$

$$\begin{aligned}
&\left[\frac{d^2}{dr^2} - \frac{\kappa(\kappa-1)}{r^2} + \frac{2\kappa}{r}U(r) + \frac{dU(r)}{dr} - U^2(r)\right]G_{n\kappa}(r) \\
&+ \frac{\frac{d\Sigma(r)}{dr}}{M - E_{n\kappa} + \Sigma(r)}\left(\frac{d}{dr} - \frac{\kappa}{r} + U(r)\right)G_{n\kappa}(r) \\
&= \left[\left(M + E_{n\kappa} - \Delta(r)\right)\left(M - E_{n\kappa} + \Sigma(r)\right)\right]G_{n\kappa}(r)
\end{aligned} \tag{11}$$

respectively, where $\kappa(\kappa-1) = \tilde{l}(\tilde{l}+1)$ and $\kappa(\kappa+1) = l(l+1)$. The quantum number $\kappa$ is related to the quantum numbers for spin symmetry $l$ and p-spin symmetry $\tilde{l}$ as

$$\kappa = \begin{cases} -(l+1) = -(j+1/2) & (s_{1/2}, p_{3/2}, etc.) \quad j = l+\frac{1}{2}, \quad \text{aligned spin} (\kappa < 0) \\ +l = +(j+1/2) & (p_{1/2}, d_{3/2}, etc.) \quad j = l-\frac{1}{2}, \quad \text{unaligned spin} (\kappa > 0), \end{cases} \quad (12)$$

and the quasidegenerate doublet structure can be expressed in terms of a p-spin angular momentum $\tilde{s} = 1/2$ and pseudo-orbital angular momentum $\tilde{l}$, which is defined as

$$\kappa = \begin{cases} -\tilde{l} = -(j+1/2) & (s_{1/2}, p_{3/2}, etc.) \quad j = \tilde{l}-\frac{1}{2}, \quad \text{aligned p-spin} (\kappa < 0) \\ +(\tilde{l}+1) = +(j+1/2) & (d_{3/2}, f_{5/2}, etc.) \quad j = \tilde{l}+\frac{1}{2}, \quad \text{unaligned p-spin} (\kappa > 0), \end{cases} \quad (13)$$

where $\kappa = \pm 1, \pm 2, \ldots$. For example, $(1s_{1/2}, 0d_{3/2})$ and $(1p_{3/2}, 0f_{5/2})$ can be considered as p-spin doublets.

## 2.1. P-spin symmetry

Ginocchio showed that there is a connection between p-spin symmetry and near equality of the time component of a vector potential and the scalar potential, $V(r) \approx -S(r)$ [7]. After that, Meng *et al.* derived that if $\frac{d[V(r)+S(r)]}{dr} = \frac{d\Sigma(r)}{dr} = 0$ or $\Sigma(r) = C_{ps} = $ constant, then p-spin symmetry is exact in the Dirac equation [43,44]. In this part, we are taking $\Delta(r)$ as the our novel potential (1) and $U(r)$ is the tensor Coulomb-like interaction, that is,

$$\Delta(r) = -\frac{A}{r^2} e^{-2\alpha r} + \frac{B}{r} e^{-\alpha r} - C, \quad (14)$$

$$U(r) = -\frac{T}{r}, \quad T = \frac{Z_a Z_b e^2}{4\pi\varepsilon_0}, \quad r \geq R_c, \quad (15)$$

where $R_c = 7.78 fm$ is the Coulomb radius, $Z_a$ and $Z_b$ denote the charges of the projectile $a$ and the target nuclei $b$, respectively [10]. Under this symmetry, Eq. (11) recasts in the simple form:

$$\left[\frac{d^2}{dr^2} - \frac{\Lambda_\kappa(\Lambda_\kappa - 1)}{r^2} + \tilde{\gamma}\left(\frac{A}{r^2}e^{-2\alpha r} - \frac{B}{r}e^{-\alpha r} + C\right) - \tilde{\beta}^2\right]G_{n\kappa}(r) = 0, \tag{16}$$

where $\kappa = -\tilde{l}$ and $\kappa = \tilde{l} + 1$ for $\kappa < 0$ and $\kappa > 0$, respectively. Also, we identified $\Lambda_\kappa = \kappa + T$, $\tilde{\gamma} = E_{n\kappa} - M - C_{ps}$ and $\tilde{\beta}^2 = (M + E_{n\kappa})(M - E_{n\kappa} + C_{ps})$.

## 2.2. Spin symmetry limit

In the spin symmetry limit $\frac{d\Delta(r)}{dr} = 0$ or $\Delta(r) = C_s = $ constant [43,44], with $\Sigma(r)$ is taken as GIQY potential (1) and $U(r)$ is the Coulomb-like tensor interaction. Thus, Eq. (10) recasts in the form:

$$\left[\frac{d^2}{dr^2} - \frac{\eta_\kappa(\eta_\kappa - 1)}{r^2} + \gamma\left(\frac{A}{r^2}e^{-2\alpha r} - \frac{B}{r}e^{-\alpha r} + C\right) - \beta^2\right]F_{n\kappa}(r) = 0, \tag{17}$$

where $\kappa = l$ and $\kappa = -l - 1$ for $\kappa < 0$ and $\kappa > 0$, respectively. Also, $\eta_\kappa = \kappa + T + 1$, $\gamma = M + E_{n\kappa} - C_s$ and $\beta^2 = (M - E_{n\kappa})(M + E_{n\kappa} - C_s)$.

Since the Dirac equation with the GIQY potential has no exact solution, we use an approximation for the centrifugal term as [20,40,45-47]

$$\frac{1}{r^2} \approx \lim_{\alpha \to 0}\left[4\alpha^2 \frac{e^{-2\alpha r}}{(1 - e^{-2\alpha r})^2}\right]. \tag{18}$$

Finally, for the solutions of Eq. (16) and Eq. (17) with the above approximation, we will employ the NU method which is briefly introduced in the following section.

## 3. Parametric NU Method

This powerful mathematical tool solves second order differential equations. Let us consider the following differential equation [49-51]

$$\psi_n''(s) + \frac{\tilde{\tau}(s)}{\sigma(s)}\psi_n'(s) + \frac{\tilde{\sigma}(s)}{\sigma^2(s)}\psi_n(s) = 0, \tag{19}$$

where $\sigma(s)$ and $\tilde{\sigma}(s)$ are polynomials, at most of second degree, and $\tilde{\tau}(s)$ is a first-degree polynomial. To make the application of the NU method simpler and direct without need to check the validity of solution. We present a shortcut for the method. So, at first we write the general form of the Schrödinger-like equation (19) in a more general form as [50,51]

$$\psi_n''(s) + \left( \frac{c_1 - c_2 s}{s(1-c_3 s)} \right) \psi_n'(s) + \left( \frac{-p_2 s^2 + p_1 s - p_0}{s^2 (1-c_3 s)^2} \right) \psi_n(s) = 0, \tag{20}$$

satisfying the wave functions

$$\psi_n(s) = \phi(s) y_n(s). \tag{21}$$

Comparing (20) with its counterpart (19), we obtain the following identifications:

$$\tilde{\tau}(s) = c_1 - c_2 s, \quad \sigma(s) = s(1-c_3 s), \quad \tilde{\sigma}(s) = -p_2 s^2 + p_1 s - p_0, \tag{22}$$

Following the NU method [49], we obtain the energy equation [50,51]

$$c_2 n - (2n+1)c_5 + (2n+1)\left(\sqrt{c_9} - c_3 \sqrt{c_8}\right) + n(n-1)c_3 + c_7 + 2c_3 c_8 - 2\sqrt{c_8 c_9} = 0, \tag{23}$$

and the wave functions

$$\rho(s) = s^{c_{10}} (1-c_3 s)^{c_{11}}, \quad \phi(s) = s^{c_{12}} (1-c_3 s)^{c_{13}}, \quad c_{12} > 0, \ c_{13} > 0,$$

$$y_n(s) = P_n^{(c_{10}, c_{11})}(1 - 2c_3 s), \quad c_{10} > -1, \ c_{11} > -1,$$

$$\psi_{n\kappa}(s) = N_{n\kappa} s^{c_{12}} (1-c_3 s)^{c_{13}} P_n^{(c_{10}, c_{11})}(1 - 2c_3 s). \tag{24}$$

where $P_n^{(\mu,\nu)}(x)$, $\mu > -1$, $\nu > -1$, and $x \in [-1,1]$ are Jacobi polynomials with the constants are

$$c_4 = \frac{1}{2}(1 - c_1), \qquad c_5 = \frac{1}{2}(c_2 - 2c_3),$$

$$c_6 = c_5^2 + p_2; \qquad c_7 = 2c_4 c_5 - p_1,$$

$$c_8 = c_4^2 + p_0, \qquad c_9 = c_3(c_7 + c_3 c_8) + c_6,$$

$$c_{10} = c_1 + 2c_4 - 2\sqrt{c_8} - 1 > -1, \qquad c_{11} = 1 - c_1 - 2c_4 + \frac{2}{c_3}\sqrt{c_9} > -1, \ c_3 \neq 0,$$

$$c_{12} = c_4 - \sqrt{c_8} > 0, \qquad c_{13} = -c_4 + \frac{1}{c_3}(\sqrt{c_9} - c_5) > 0, \ c_3 \neq 0, \tag{25}$$

where $c_{12} > 0$, $c_{13} > 0$ and $s \in [0, 1/c_3]$, $c_3 \neq 0$.

In the rather more special case of $c_3 = 0$, the wave function (21) becomes

$$\lim_{c_3 \to 0} P_n^{(c_{10}, c_{11})}(1 - 2c_3 s) = L_n^{c_{10}}(c_{11} s), \quad \lim_{c_3 \to 0} (1-c_3 s)^{c_{13}} = e^{c_{13} s},$$

$$\psi(s) = N s^{c_{12}} e^{c_{13} s} L_n^{c_{10}}(c_{11} s). \tag{26}$$

## 4. Bound State Solutions of the Dirac Equation

We will solve the Dirac equation with our novel potential including tensor interaction by using the NU method.

**4.1. P-spin symmetric case**

After inserting approximation (18) into Eq. (16), we obtain the following form which is solvable by NU method:

$$\left[\frac{d^2}{dr^2} - 4\alpha^2 \Lambda_\kappa(\Lambda_\kappa - 1)\frac{e^{-2\alpha r}}{(1-e^{-2\alpha r})^2} + \tilde{\gamma}\left(4\alpha^2 A \frac{e^{-4\alpha r}}{(1-e^{-2\alpha r})^2} - 2\alpha B \frac{e^{-2\alpha r}}{1-e^{-2\alpha r}} + C\right) - \tilde{\beta}^2\right]$$
$$\times G_{n\kappa}(r) = 0,$$

and further making transformation of variables $s = e^{-2\alpha r}$, mapping $r \in (0,\infty)$ into $s \in (0,1)$, we can rewrite it as follows

$$\left[\frac{d^2}{ds^2} + \frac{1-s}{s(1-s)}\frac{d}{ds} + \frac{1}{s^2(1-s)^2}\right.$$
$$\left. \times \left(-\Lambda'_\kappa(\Lambda'_\kappa - 1)s + \tilde{\gamma}'s^2 - \frac{\tilde{\beta}'^2}{4\alpha^2}(1-s)^2\right)\right] G_{n\kappa}(r) = 0, \quad (27)$$

with

$$\Lambda'_\kappa(\Lambda'_\kappa - 1) = \Lambda_\kappa(\Lambda_\kappa - 1) + \frac{\tilde{\gamma}B}{2\alpha}, \quad \tilde{\gamma}' = \tilde{\gamma}\left(\frac{B+2\alpha A}{2\alpha}\right), \quad \tilde{\beta}'^2 = \tilde{\beta}^2 - \tilde{\gamma}C. \quad (28)$$

Comparing Eq. (27) with Eq. (20), we get

$$c_1 = 1, \qquad p_2 = \frac{\tilde{\beta}'^2}{4\alpha^2} - \tilde{\gamma}',$$
$$c_2 = 1, \qquad p_1 = -\Lambda'_\kappa(\Lambda'_\kappa - 1) + \frac{\tilde{\beta}'^2}{2\alpha^2}, \quad (29)$$
$$c_3 = 1, \qquad p_0 = \frac{\tilde{\beta}'^2}{4\alpha^2},$$

and by the use of Eq. (25), we can obtain

$$c_4 = 0, \qquad c_5 = -\frac{1}{2},$$

$$c_6 = \frac{1}{4} + \frac{\tilde{\beta}'^2}{4\alpha^2} - \tilde{\gamma}', \qquad c_7 = \Lambda'_\kappa(\Lambda'_\kappa - 1) - \frac{\tilde{\beta}'^2}{2\alpha^2},$$

$$c_8 = \frac{\tilde{\beta}'^2}{4\alpha^2}, \qquad c_9 = \lambda^2, \qquad (30)$$

$$c_{10} = -\frac{\tilde{\beta}'}{\alpha}, \qquad c_{11} = 2\lambda,$$

$$c_{12} = -\frac{\tilde{\beta}'}{2\alpha}, \qquad c_{13} = \frac{1}{2} + \lambda,$$

where $\lambda = \sqrt{(\Lambda'_\kappa - 1/2)^2 - \tilde{\gamma}'}$. The energy equation can be obtained by using Eqs. (23), (29) and (30) to get

$$\left(n + \frac{1}{2} + \sqrt{(\Lambda'_\kappa - 1/2)^2 - \tilde{\gamma}'} - \frac{\tilde{\beta}'}{2\alpha}\right)^2 = \frac{\tilde{\beta}'^2}{4\alpha^2} - \tilde{\gamma}', \qquad (31)$$

with

$$\Lambda'_\kappa = \frac{1}{2} + \sqrt{(\kappa + T - 1/2)^2 + \frac{B}{2\alpha}(E_{n\kappa} - M - C_{ps})},$$

$$\tilde{\gamma}' = (E_{n\kappa} - M - C_{ps})\left(\frac{B + 2\alpha A}{2\alpha}\right),$$

$$\tilde{\beta}' = \sqrt{(M - E_{n\kappa} + C_{ps})(M + E_{n\kappa} + C)}. \qquad (32)$$

It is worthy to note that once $B = C = 0$, Eq. (31) reduces to Eq. (30) of Ref. [20] giving the identical energy spectra presented in Tables 1 and 2 of Ref. [20].

Further, substituting the explicit forms of $\Lambda'_\kappa$, $\tilde{\gamma}'$ and $\tilde{\beta}'^2$ of Eq. (32) into Eq. (31), one can readily obtain the energy formula in closed form. In the limiting case when the screening parameter $\alpha \to 0$ (low screening regime), the potential approximates as

$V_{IQY}(r) = -V_0 \lim_{\alpha \to 0}\left(1 - \frac{1}{r}e^{-\alpha r}\right)^2 \simeq \frac{a}{r^2} - \frac{b}{r} + c$, where the potential parameters are defined as $a = -V_0$, $b = -2(1+\alpha)V_0$, $c = -(1+\alpha)^2 V_0$. This potential is well known as Mie-type potential [18, 27]. The energy eigenvalue equation for this potential has recently been found in Ref. [27] as

$$\sqrt{(E_{n\kappa} - M - C_{ps})c + (M + E_{n\kappa})(M - E_{n\kappa} + C_{ps})}$$

$$= \frac{(E_{n\kappa} - M - C_{ps})b}{1 + 2n + 2\sqrt{(\kappa - 1/2)^2 + (E_{n\kappa} - M - C_{ps})a}}. \quad (33)$$

In the special case when $a = c = 0$ and $C_{ps} = 0$, it gives the energy formula for the Coulomb-like potential [27,51]

$$E_{n\kappa} = -M \frac{4(n+\kappa)^2 - b^2}{4(n+\kappa)^2 + b^2}. \quad (34)$$

Furthermore, when $n \to \infty$, one obtains for all states $E_{n\kappa} = -M$ (continuum states), that is, it shows that when $n$ goes to infinity the energy solution of Eq. (31) becomes finite (this is the exact p-spin symmetric case given by Eq. (38) of Ref. [51]).

On the other hand, to find the corresponding wave functions, referring to Eq. (24), we find the functions

$$\rho(s) = s^{-\tilde{\beta}'/\alpha}(1-s)^{2\lambda}, \quad (35)$$

$$\phi(s) = s^{(-\tilde{\beta}'/2\alpha)}(1-s)^{\frac{1}{2}+\lambda}. \quad (36)$$

Hence, Eq. (24) with the help of the weight function (35) gives

$$y_n(s) = P_n^{(-\tilde{\beta}'/\alpha, 2\lambda)}(1-2s). \quad (37)$$

By using $G_{n\kappa}(s) = \phi(s) y_n(s)$, we obtain the lower spinor component of the wave function as

$$G_{n\kappa}(r) = A_{n\kappa}\left(e^{-2\alpha r}\right)^{(-\tilde{\beta}'/2\alpha)}(1-e^{-2\alpha r})^{\frac{1}{2}+\lambda} P_n^{(-\tilde{\beta}'/\alpha, 2\lambda)}\left(1 - 2e^{-2\alpha r}\right)$$

$$= A'_{n\kappa}\left(e^{-2\alpha r}\right)^{(-\tilde{\beta}'/2\alpha)}(1-e^{-2\alpha r})^{\frac{1}{2}+\lambda}{}_2F_1\left(-n, n - \tilde{\beta}'/\alpha + 2\lambda + 1, -\tilde{\beta}'/\alpha + 1; e^{-2\alpha r}\right), \quad (38)$$

$$A'_{n\kappa} = \frac{\Gamma(n + \tilde{\beta}'/\alpha + 1)}{\Gamma(\tilde{\beta}'/\alpha + 1)n!} A_{n\kappa},$$

where $A_{n\kappa}$ is the normalization constant. The upper component of the Dirac spinor can be calculated from Eq. (8b) as [52]

$$F_{n\kappa}(r) = \frac{1}{(M - E_{n\kappa} + C_{ps})}\left\{\left[\alpha(1+2\lambda)\frac{e^{-2\alpha r}}{(1-e^{-2\alpha r})} + \tilde{\beta}' - \frac{\kappa + T}{r}\right]G_{n\kappa}(r)\right.$$

$$+ A'_{n\kappa} 2\alpha n \frac{(n - \tilde{\beta}'/\alpha + 2\lambda + 1)}{(-\tilde{\beta}'/\alpha + 1)}\left(e^{-2\alpha r}\right)^{(-\tilde{\beta}'/2\alpha)+1}(1-e^{-2\alpha r})^{\frac{1}{2}+\lambda} \quad (39)$$

$$\left. \times {}_2F_1\left(-n+1, n - \tilde{\beta}'/\alpha + 2\lambda + 2, -\tilde{\beta}'/\alpha + 2; e^{-2\alpha r}\right)\right\}.$$

and for the finiteness of our solution we demand that $E_{n\kappa} \neq M + C_{ps}$. Further, the exact p-spin symmetric case occurs when $C_{ps} = 0$, which gives negative energy eigenvalues. The finiteness of our solution requires that the two-components of the wave function be well defined over the whole positive range, $r \in (0, \infty)$. However, in the p-spin limit, if the positive energy is chosen, the upper-spinor component of the wave function will be no longer defined as obviously noted in Eq. (39). Note that the introduction of the Coulomb-like tensor interaction has no much effect on the negativity of the energy spectrum in the p-spin limit, but the main purpose is just the tendency to remove the degeneracy in the p-spin doublets.

Of course, the energy eigenvalue equation (31) admits two solutions (negative and positive), however, we are obliged to choose the negative energy solution to make the wave function normalizable (integrable) in the half positive line range [52-54].

**4.2. Spin symmetric case**

Inserting approximation (18) into Eq. (17), we obtain a simple form which is amendable by the solution of NU method:

$$\left[\frac{d^2}{dr^2} - 4\alpha^2 \eta_\kappa (\eta_\kappa - 1) \frac{e^{-2\alpha r}}{(1-e^{-2\alpha r})^2} + \gamma \left( 4\alpha^2 A \frac{e^{-4\alpha r}}{(1-e^{-2\alpha r})^2} - 2\alpha B \frac{e^{-2\alpha r}}{1-e^{-2\alpha r}} + C \right) - \beta^2 \right]$$
$$\times F_{n\kappa}(r) = 0.$$

To avoid repetition in the solution of Eq. (17),

$$\left[\frac{d^2}{ds^2} + \frac{1-s}{s(1-s)} \frac{d}{ds} + \frac{1}{s^2(1-s)^2} \right.$$
$$\left. \times \left( -\eta'_\kappa (\eta'_\kappa - 1)s + \gamma' V_0 s^2 - \frac{\beta'^2}{4\alpha^2}(1-s)^2 \right) \right] F_{n\kappa}(r) = 0, \quad (40)$$

$$\eta'_\kappa = \frac{1}{2} + \sqrt{(\kappa + T + 1/2)^2 + \frac{B}{2\alpha}(M + E_{n\kappa} - C_s)},$$

$$\gamma' = \left(\frac{B + 2\alpha A}{2\alpha}\right)(M + E_{n\kappa} - C_s),$$

$$\beta' = \sqrt{(M + E_{n\kappa} - C_s)(M - E_{n\kappa} - C)}. \quad (41)$$

We follow the same procedures explained in the subsection 4.1 and hence obtain the following energy eigenvalue equation using the parametric mapping transformation (cf. Eq. (38) of Ref. [39] and Eq. (49) of Ref. [52]):

$$\left(n+\frac{1}{2}+\sqrt{(\eta'_\kappa-1/2)^2-\gamma'}-\frac{\beta'}{2\alpha}\right)^2=\frac{\beta'^2}{4\alpha^2}-\gamma', \tag{42}$$

and the corresponding wave functions for the upper Dirac spinor as

$$F_{n\kappa}(r)=B_{n\kappa}e^{-(\beta'/2\alpha)r}(1-e^{-2\alpha r})^{\frac{1}{2}+\sigma}P_n^{(-\beta'/\alpha,2\sigma)}\left(1-2e^{-2\alpha r}\right), \tag{43}$$

where $B_{nk}$ is the normalization constant and $\sigma=\sqrt{(\eta'_\kappa-1/2)^2-\gamma'}$. Finally, the lower-spinor component of the Dirac equation can be obtained via Eq. (8a) as

$$G_{n\kappa}(r)=\frac{1}{(M+E_{n\kappa}-C_s)}\left\{\left[\alpha(1+2\sigma)\frac{e^{-2\alpha r}}{(1-e^{-2\alpha r})}+\beta'+\frac{\kappa+T}{r}\right]F_{n\kappa}(r)\right.$$

$$+A'_{n\kappa}2\alpha n\frac{(n-\beta'/\alpha+2\sigma+1)}{(\beta'/\alpha+1)}\left(e^{-2\alpha r}\right)^{(-\beta'/2\alpha)+1}(1-e^{-2\alpha r})^{\frac{1}{2}+\sigma} \tag{44}$$

$$\left.\times{}_2F_1\left(-n+1,n-\beta'/\alpha+2\sigma+2,-\beta'/\alpha+2;e^{-2\alpha r}\right)\right\}.$$

where $E_{n\kappa}\neq-M+C_s$.

In the non-relativistic limit, the energy levels can be obtained when $C_s=0$, $T=0$, $E_{n\kappa}+M\approx2\mu/\hbar^2$, $E_{n\kappa}-M\approx E_{nl}$ and $\kappa(\kappa+1)\to l(l+1)$ from (42) as

$$E_{nl}=-C-\frac{2\mu}{\hbar^2}$$

$$\times\left[\frac{\alpha\left(n+1/2+\sqrt{(l+1/2)^2-2\mu A/\hbar^2}\right)\hbar^2}{2\mu}+\frac{(B+2\alpha A)}{2\left(n+1/2+\sqrt{(l+1/2)^2-2\mu A/\hbar^2}\right)}\right]^2, \tag{45}$$

and the radial wave functions from Eq. (43) are

$$R_{nl}(r)=D_{nl}e^{-\sqrt{-2\mu(E_{nl}+C)/\hbar^2}\,r}(1-e^{-2\alpha r})^{1/2+\sqrt{(l+1/2)^2-2\mu A/\hbar^2}}$$

$$\times P_n^{\left(\sqrt{-2\mu(E_{nl}+C)/\hbar^2}/\alpha,2\sqrt{(l+1/2)^2-2\mu A/\hbar^2}\right)}\left(1-2e^{-2\alpha r}\right), \tag{46}$$

When $B=C=0$, we obtain the non-relativistic energy levels of the IQY potential:

$$E_{nl} = -\frac{2\mu}{\hbar^2}$$

$$\times \left[ \frac{\alpha\left(n+1/2+\sqrt{(l+1/2)^2-2\mu A/\hbar^2}\right)\hbar^2}{2\mu} + \frac{2\alpha A}{2\left(n+1/2+\sqrt{(l+1/2)^2-2\mu A/\hbar^2}\right)} \right]^2, \quad (47)$$

and the corresponding wave functions

$$R_{nl}(r) = D_{nl} e^{-\sqrt{-2\mu E_{nl}/\hbar^2}\, r} (1-e^{-2\alpha r})^{1/2+\sqrt{(l+1/2)^2-2\mu A/\hbar^2}}$$

$$\times P_n^{\left(\sqrt{-2\mu E_{nl}/\hbar^2}/\alpha,\, 2\sqrt{(l+1/2)^2-2\mu A/\hbar^2}\right)} (1-2e^{-2\alpha r}), \quad (48)$$

Further, when $A = C = 0$, and $B \to -A$, then it turns to be the Yukawa potential with the energy formula obtained via Eq. (45) becomes

$$E_{nl} = -\frac{2\mu}{\hbar^2} \left[ \frac{A}{2(n+l+1)} - \frac{\hbar^2(n+l+1)}{2\mu}\alpha \right]^2, \quad (49)$$

which is identical to Eq. (52) of Ref. [40] and the corresponding wave functions

$$R_{nl}(r) = D_{nl} e^{-\sqrt{-2\mu E_{nl}/\hbar^2}\, r} (1-e^{-2\alpha r})^{l+1} P_n^{\left(\sqrt{-2\mu E_{nl}/\hbar^2}/\alpha,\, 2l+1\right)} (1-2e^{-2\alpha r}), \quad (50)$$

where the normalization constant $D_{nl}$ is to be found by the normalization condition.

### 4.3. Some remarks and numerical results

It is worthy to note that the tensor interaction generates a new spin-orbit centrifugal term $\Lambda(\Lambda \pm 1)$ where $\Lambda = \Lambda'_\kappa$ or $\eta'_\kappa$. Some numerical results are given in table 1 and table 2, where we use the parameter values as $M = 5.0\, fm^{-1}$, $V_0 = 1.0$, $C_{ps} = -6\, fm^{-1}$ and $C_s = 6.0\, fm^{-1}$. In table 1, we consider the same set of p-spin symmetry doublets: $(1s_{1/2}, 0d_{3/2})$, $(1p_{3/2}, 0f_{5/2})$, $(1d_{5/2}, 0g_{7/2})$, $(1f_{7/2}, 0h_{9/2})$, …. In table 2, we consider the same set of spin symmetry doublets: $(0p_{1/2}, 0p_{3/2})$, $(0d_{3/2}, 0d_{5/2})$, $(0f_{5/2}, 0f_{7/2})$, $(0g_{7/2}, 0g_{9/2})$, …. We have noticed that the tensor interaction removes the degeneracy between two states in spin doublets and p-spin doublets. When $T \neq 0$, the energy levels of the spin (p-spin) aligned states and spin (p-spin) unaligned states move in the opposite directions. For example, in p-spin doublet $(1s_{1/2}, 0d_{3/2})$; when $T = 0$, $E_{1,-1} = E_{1,2} = -1.000402620\, fm^{-1}$, but when $T = 0.5$, $E_{1,-1} = -1.000279597\, fm^{-1}$ with $\kappa < 0$ and $E_{1,2} = -1.000548021\, fm^{-1}$ with $\kappa > 0$. Also, we compared our results with the usual Yukawa potential.

## 5. Conclusion

In this paper, we have introduced a novel potential which is an intermediate between the Yukawa potential [40] and the inverse quadratic Yukawa (IQY) potential [20] that milds the strong singularity of $1/r^2$ in IQY potential and the soft singularity $1/r$ in Yukawa potential. As shown in Figure 1, the behavior of the novel Yukawa-like potential is very close and overlapping with the other two potentials studied before by us in [20,40]. It can be easily reduced into the other potential forms [20,40]. We have obtained the approximate bound states of a Dirac particle confined to the field the GIQY interaction in the presence of spin and p-spin symmetry including Coulomb-like tensor interaction in the form of $-T/r$. We used a parametric version of the powerful NU method [49]. Some numerical values of the energy levels are calculated in tables 1 and 2 in view of p-spin and spin symmetry, respectively. Obviously, the degeneracy between the members of doublet states in p-spin and spin symmetries is removed by tensor interaction. The spin and p-spin spectra of the present potential is identical to those ones obtained in [20] as the potential parameters $B = C = 0$. Further, it has a continuum spectrum once $n \to \infty$. Finally, the relativistic spin symmetry in the absence of tensor interaction ($T = 0$) and when $C_s = 0$, reduces into the Schrödinger solution for the Yukawa potential [40] and the IQY potential [20] under appropriate transformations mapping of parameters [39,52].

**Acknowledgments**

We thank the referees and editor for invaluable suggestions which have greatly helped in improving the manuscript. S.M. Ikhdair acknowledges the partial support of the Scientific and Technological Research Council of Turkey.

**References**

[1] J. N. Ginocchio, Relativistic symmetries in nuclei and hadrons, Phys. Rep. **414** (4-5) (2005) 165.

[2] A. Bohr, I. Hamamoto and B. R. Mottelson, Pseudospin in rotating nuclear potentials, Phys. Scr. **26** (1982) 267.

[3] J. Dudek, W. Nazarewicz, Z. Szymanski and G. A. Leander, Abundance and systematics of nuclear superdeformed states; relation to the pseudospin and pseudo-SU(3) symmetries, Phys. Rev. Lett. **59** (1987) 1405.

[4] D. Troltenier, C. Bahri and J. P. Draayer, Generalized pseudo-SU(3) model and pairing, Nucl. Phys. A **586** (1995) 53.

[5] P. R. Page, T. Goldman and J. N. Ginocchio, Relativistic symmetry suppresses quark spin-orbit splitting, Phys. Rev. Lett. **86** (2001) 204.

[6] J. N. Ginocchio, A. Leviatan, J. Meng, and S. G. Zhou, Test of pseudospin symmetry in deformed nuclei, Phys. Rev. C **69** (2004) 034303.

[7] J. N. Ginocchio, Pseudospin as a relativistic symmetry, Phys. Rev. Lett. **78** (3) (1997) 436.

[8] K. T. Hecht and A. Adler, Generalized seniority for favored $J \neq 0$ pairs in mixed configurations, Nucl. Phys. A **137** (1969) 129.

[9] A. Arima, M. Harvey and K. Shimizu, Pseudo LS coupling and pseudo SU3 coupling schemes, Phys. Lett. B **30** (1969) 517.

[10] S. M. Ikhdair and R. Sever, Approximate bound state solutions of Dirac equation with Hulthén potential including Coulomb-like tensor potential, Appl. Math. Com. **216** (2010) 911.

[11] M. Moshinsky and A. Szczepanika, The Dirac oscillator, J. Phys. A: Math. Gen. **22** (1989) L817.

[12] V. I. Kukulin, G. Loyla and M. Moshinsky, A Dirac equation with an oscillator potential and spin-orbit coupling, Phys. Lett. A **158** (1991) 19.

[13] R. Lisboa, M. Malheiro, A. S. de Castro, P. Alberto and M. Fiolhais, Pseudospin symmetry and the relativistic harmonic oscillator, Phys. Rev. C **69** (2004) 024319.

[14] P. Alberto, R. Lisboa, M. Malheiro and A. S. de Castro, Tensor coupling and pseudospin symmetry in nuclei, Phys. Rev. C **71** (2005) 034313.

[15] H. Akçay, Dirac equation with scalar and vector quadratic potentials and Coulomb-like tensor potential, Phys. Lett. A **373** (2009) 616.

[16] H. Akçay, The Dirac oscillator with a Coulomb-like tensor potential, J. Phys. A: Math. Theor. **40** (2007) 6427.

[17] O. Aydoğdu and R. Sever, Exact pseudospin symmetric solution of the Dirac equation for pseudoharmonic potential in the presence of tensor potential, Few-Body Syst. **47** (2010) 193.

[18] M. Hamzavi, A. A. Rajabi and H. Hassanabadi, Exact spin and pseudospin symmetry solutions of the Dirac equation for Mie-type potential including a Coulomb-like tensor potential, Few-Body Syst. **48** (2010) 171.

[19] M. Hamzavi, A. A. Rajabi and H. Hassanabadi, Relativistic Morse potential and tensor interaction, Few-Body Syst. **52** (2012) 19.

[20] M. Hamzavi, S. M. Ikhdair and B. I. Ita, Approximate spin and pseudospin solutions to the Dirac equation for the inversely quadraticYukawa potential and tensor interaction, Phys. Scr. **85** (2012) 045009.

[21] R. Sever, C. Tezcan, M. Aktaş and O. Yeşiltaş, Exact solution of Schrödinger equation for pseudoharmonic potential, J. Math. Chem. **43** (2007) 845.

[22] S. M. Ikhdair and R. Sever, Exact polynomial eigensolutions of the Schrödinger equation for the pseudoharmonic potential, J. Mol. Struct.: Theochem **806** (2007) 155.

[23] S. H. Dong, X. Y. Gu, Z. Q. Ma and S. Dong, Exact solutions of the Dirac equation with a Coulomb plus scalar potential in 2+1 dimensions, Int. J. Mod. Phys. E **11** (2002) 483.

[24] D. G. C. McKeon and G. V. Leeuwen, The Dirac equation in a pseudoscalar Coulomb potential, Mod. Phys. Lett. A **17** (2002) 1961.

[25] S. M. Ikhdair and R. Sever, Exact solutions of the radial Schrödinger equation for some physical potentials, Cent. Eur. J. Phys. **5** (2007) 516.

[26] M. Hamzavi, H. Hassanabadi and A. A. Rajabi, Exact solutions of Dirac equation with Hartmannn potential by Nikiforov-Uvarov method, Int. J. Mod. Phys. E **19** (2010) 2189.

[27] M. Hamzavi, H. Hassanabadi and A. A. Rajabi, Mod. Exact solution of Dirac equation for Mie-type potential by using the Nikiforov-Uvarov method under the pseudospin and spin symmetry limit, Phys. Lett. A **25** (2010) 2447.

[28] M. Hamzavi, A. A. Rajabi and H. Hassanabadi, Exact pseudospin symmetry solution of the Dirac equation for spatially-dependent mass Coulomb potential including a Coulomb-like tensor interaction via asymptotic iteration method, Phys. Lett. A **374** (2010) 4303.

[29] L. H. Zhang, X. P. Li and C. S. Jia, Approximate analytical solutions of the Dirac equation with the generalized Morse potential model in the presence of the spin symmetry and pseudo-spin symmetry, Phys. Scr. **80** (2009) 035003.

[30] S. H. Dong and M. Lozada-Cassou, On the analysis of the eigenvalues of the Dirac equation with a $1/r$ potential in $D$ dimensions, Int. J. Mod. Phys. E **13** (2004) 917.

[31] S. M. Ikhdair and R. Sever, Exact bound states of the D-dimensional Klein-Gordon equation with equal scalar and vector ring-shaped pseudoharmonic potentials, Int. J. Mod. Phys. C **19** (2008) 1425.

[32] S. M. Ikhdair, Effective Schrödinger equation with general ordering ambiguity position-dependent mass Morse potential, Mol. Phys. **110** (2012) 1415.

[33] S.M. Ikhdair and R. Sever, Two approximation schemes to the bound states of the Dirac-Hulthen problem, J. Phys. A: Math. Theor. **44** (2011) 345301-29

[34] S. H. Dong, The realization of dynamic group for the pseudoharmonic oscillator, Appl. Math. Lett. **16** (2003) 199.

[35] M. C. Zhang, G. Q. Huang-Fu and B. An, Pseudospin symmetry for a new ring-shaped non-spherical harmonic oscillator potential, Phys. Scr. **80** (2009) 065018.

[36] G. F. Wei and S. H. Dong, Pseudospin symmetry in the relativistic Manning-Rosen potential including a Pekeris-type approximation to the pseudo-centrifugal term, Phys. Lett. B **686** (2010) 288.

[37] S. M. Ikhdair, On the bound-state solutions of the Manning-Rosen potential including an improved approximation to the orbital centrifugal term, Phys. Scr. **83** (2011) 015010.

[38] K. J. Oyewumi, F. O. Akinpelu and A. D. Agboola, Exactly Complete Solutions of the Pseudoharmonic Potential in N-Dimensions, Int. J. Theor. Phys. **47** (2008) 1039.

[39] S. M. Ikhdair and R. Sever, Approximate bound states of the Dirac equation with some physical quantum potentials, Appl. Math. Comput. 218 (2012) 10082.

[40] S. M. Ikhdair, Approximate $\kappa$-state solutions to the Dirac-Yukawa problem based on the spin and pseudospin symmetry, Cent. Eur. J. Phys. 10 (2012) 361.

[41] H. Taseli, Int. J. Quant. Chem. **63** (1997) 949.

[42] M. W. Kermode, M. L. J. Allen, J. P. Mctavish and A. Kervell, J. Phys. G: Nucl. Part. Phys. **10** (1984) 773.

[43] J. N. Ginocchio, The relativistic foundations of pseudospin symmetry in nuclei, Nucl. Phys. A **654** (1999) 663.

[44] J. N. Ginocchio, A relativistic symmetry in nuclei, Phys. Rep. **315** (1999) 231.

[45] R. L. Greene and C. Aldrich, Variational wave functions for a screened Coulomb potential, Phys. Rev. A **14** (1976) 2363.

[46] O. Aydoğdu and R. Sever, The Dirac-Yukawa problem in view of pseudospin symmetry, Phys. Scr. **84** (2011) 025005.

[47] M. R. Setare and S. Haidari, Spin symmetry of the Dirac equation with the Yukawa potential, Phys. Scr. **81** (2010) 065201.

[48] A. F. Nikiforov and V. B. Uvarov, Special Functions of Mathematical Physics, (Birkhausr, Berlin, 1988).

[49] S. M. Ikhdair, Rotational and vibrational diatomic molecules in the Klein-Gordon equation with hyperbolic scalar and vector potentials, Int. J. Mod. Phys. C **20** (2009) 1563.

[50] C. Tezcan and R. Sever, A general approach for the exact solution of the Schrödinger equation, Int. J. Theor. Phys. **48** (2009) 337.

[51] S. M. Ikhdair, Exact Klein-Gordon equation with spatially-dependent masses for unequal scalar-vector Coulomb-like potentials, Eur. Phys. J. A **40** (2009) 143.

[52] S. M. Ikhdair, An approximate $\kappa$-state solutions of the Dirac equation for the generalized Morse potential under spin and pseudospin symmetry, J. Math. Phys. **52** (2011) 052303.

[53] S. M. Ikhdair and R. Sever, Relativistic and nonrelativistic bound states of the isotonic oscillator by Nikiforov-Uvarov method, J. Math. Phys. **52** (2011) 122108.

[54] S. M. Ikhdair, Approximate solutions of the Dirac equation for the Rosen-Morse potential including the spin-orbit centrifugal term, J. Math. Phys. **51** (2010) 023525.

**Table 1.** The pspin symmetric bound state energy levels (in unit of $fm^{-1}$) of the GIQY potential taking various values of $n$ and $\kappa$.

| $\tilde{l}$ | $n, \kappa < 0$ | $(l,j)$ | $E_{n,\kappa<0}$ $T=0.5$ **GIQY** | $E_{n,\kappa<0}$ $T=0$ **GIQY** | $E_{n,\kappa<0}$ $T=0$ **Yukawa** | $n-1, \kappa > 0$ | $(l+2, j+1)$ | $E_{n-1,\kappa>0}$ $T=0.5$ **GIQY** | $E_{n-1,\kappa>0}$ $T=0$ **GIQY** | $E_{n,\kappa<0}$ $T=0$ **Yukawa** |
|---|---|---|---|---|---|---|---|---|---|---|
| 1 | 1, -1 | $1s_{1/2}$ | -1.000279597 | -1.000402620 | -1.000504424 | 0, 2 | $0d_{3/2}$ | -1.000548021 | -1.000402620 | -1.000504424 |
| 2 | 1, -2 | $1p_{3/2}$ | -1.000548021 | -1.000715801 | -1.000896841 | 0, 3 | $0f_{5/2}$ | -1.000905966 | -1.000715801 | -1.000896841 |
| 3 | 1, -3 | $1d_{5/2}$ | -1.000905966 | -1.001118521 | -1.001401490 | 0, 4 | $0g_{7/2}$ | -1.001353470 | -1.001118521 | -1.001401490 |
| 4 | 1, -4 | $1f_{7/2}$ | -1.001353470 | -1.001610821 | -1.002018456 | 0, 5 | $0h_{9/2}$ | -1.001890579 | -1.001610821 | -1.002018456 |
| 1 | 2, -1 | $2s_{1/2}$ | -1.000548064 | -1.000715836 | -1.000896841 | 1, 2 | $1d_{3/2}$ | -1.000905997 | -1.000715836 | -1.000896841 |
| 2 | 2, -2 | $2p_{3/2}$ | -1.000905997 | -1.001118549 | -1.001401490 | 1, 3 | $1d_{3/2}$ | -1.001353498 | -1.001118549 | -1.001401490 |
| 3 | 2, -3 | $2d_{5/2}$ | -1.001353498 | -1.001610848 | -1.002018456 | 1, 4 | $1g_{7/2}$ | -1.001890607 | -1.001610848 | -1.002018456 |
| *4* | 2, -4 | $2f_{7/2}$ | -1.001890607 | -1.002192781 | -1.002747843 | 1, 5 | $1h_{9/2}$ | -1.002517380 | -1.002192781 | -1.002747843 |

**Table 2.** The spin symmetric bound state energy levels (in unit of $fm^{-1}$) of the GIQY potential taking various values of $n$ and $\kappa$.

| $l$ | $n, \kappa < 0$ | $(l, j = l+1/2)$ | $E_{n,\kappa<0}$ $T=0.5$ GIQY | $E_{n,\kappa<0}$ $T=0$ GIQY | $E_{n,\kappa<0}$ $T=0$ Yukawa | $n, \kappa > 0$ | $(l, j = l-1/2)$ | $E_{n,\kappa>0}$ $T=0.5$ GIQY | $E_{n,\kappa>0}$ $T=0$ GIQY | $E_{n,\kappa<0}$ $T=0$ Yukawa |
|---|---|---|---|---|---|---|---|---|---|---|
| 1 | 0, -2 | $0p_{3/2}$ | 1.000170475, 1.959485534 | 1.000303085, 2.744284589 | 1.000225861, 4.750362375 | 0, 1 | $0p_{1/2}$ | 1.000473600, 3.263495375 | 1.000303085, 2.744284589 | 1.000225861, 4.750362375 |
| 2 | 0, -3 | $0d_{5/2}$ | 1.000473600, 3.263495375 | 1.000682034, 3.527757918 | 1.000508222, 4.876789075 | 0, 2 | $0d_{3/2}$ | 1.000928404, 3.666854292 | 1.000682034, 3.527757918 | 1.000508222, 4.876789075 |
| 3 | 0, -4 | $0f_{7/2}$ | 1.000928404, 3.666854292 | 1.001212727, 3.747925241 | 1.000903596, 4.922788712 | 0, 3 | $0f_{5/2}$ | 1.001535026, 3.799364444 | 1.001212727, 3.747925241 | 1.000903596, 4.922788712 |
| 4 | 0, -5 | $0g_{9/2}$ | 1.001535026, 3.799364444 | 1.001895326, 3.834121095 | 1.001412049, 4.944132506 | 0, 4 | $0g_{7/2}$ | 1.002293655, 3.858727690 | 1.001895326, 3.834121095 | 1.001412049, 4.944132506 |
| 1 | 1, -2 | $1p_{3/2}$ | 1.000473555, ---------------- | 1.000681993, 3.201886745 | 1.000903596, 4.922788712 | 1, 1 | $1p_{1/2}$ | 1.000928362, 3.597525475 | 1.000681993, 3.201886745 | 1.000903596, 4.922788712 |
| 2 | 1, -3 | $1d_{5/2}$ | 1.000928362, 3.597525475 | 1.001212685, 3.722793657 | 1.001412049, 4.944132506 | 1, 2 | $1d_{3/2}$ | 1.001534983, 3.788057103 | 1.001212685, 3.722793657 | 1.001412049, 4.944132506 |
| 3 | 1, -4 | $1f_{7/2}$ | 1.001534983, 3.788057103 | 1.001895281, 3.828338469 | 1.002033667, 4.955483574 | 1, 3 | $1f_{5/2}$ | 1.002293607, 3.855501078 | 1.001895281, 3.828338469 | 1.002033667, 4.955483574 |
| 4 | 1, -5 | $1g_{9/2}$ | 1.002293607, 3.855501078 | 1.002729992, 3.874855744 | 1.002768558, 4.962003015 | 1, 4 | $1g_{7/2}$ | 1.003204469, 3.889177136 | 1.002729992, 3.874855744 | 1.002768558, 4.962003015 |

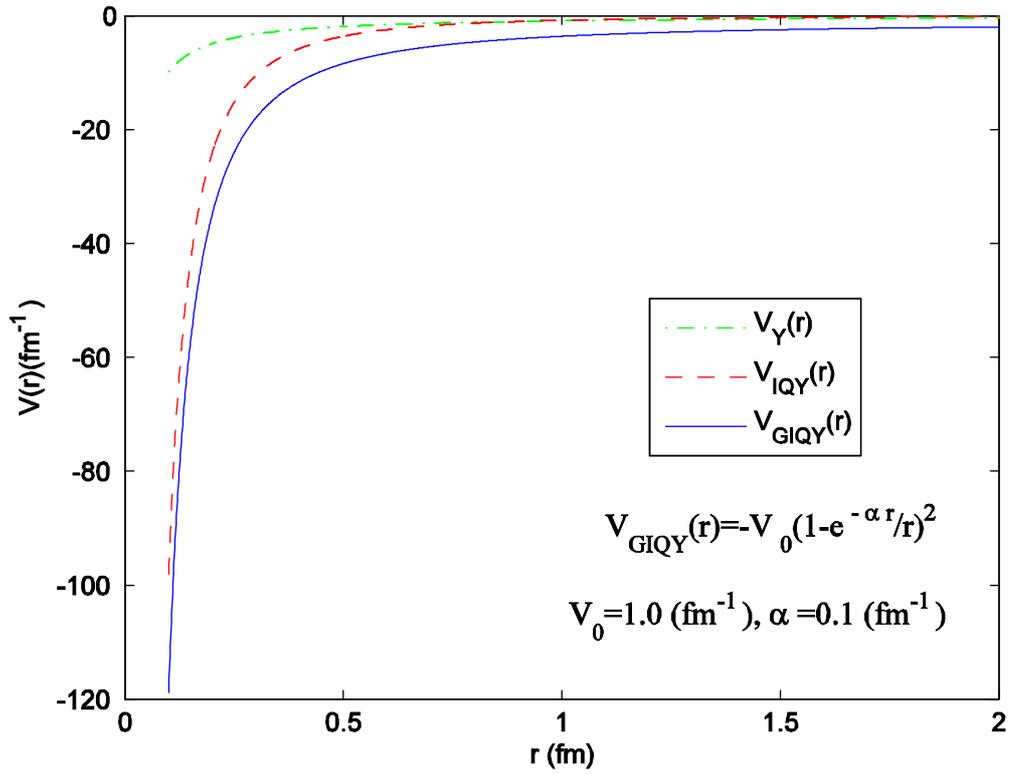

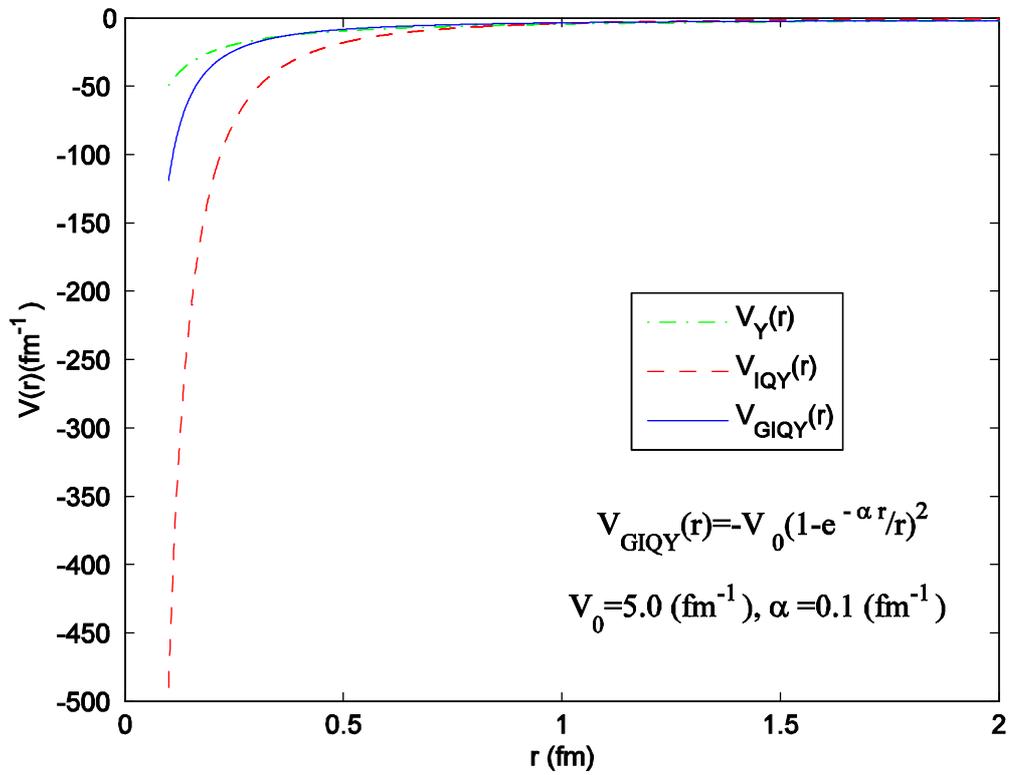

Figure 1. The behavior of the GIQY potential compared with Yukawa potential [40] and IQY potential [20] for (a) weak $V_0 = 1.0\, fm^{-1}$ and (b) strong $V_0 = 5.0\, fm^{-1}$ values for the coupling potential parameter.